\documentclass[aps,preprint,showkeys, amsfonts,showpacs,amsmath,amssymb]{revtex4-1}
\usepackage[dvips]{color}
\usepackage{epsfig}
\usepackage{amsmath}
\usepackage{graphicx}
\begin{document}
\title{Generalized Second Law of Thermodynamics with Corrected Entropy in Tachyon Cosmology  }

\author{$^1,^2$H. Farajollahi}
\email{hosseinf@guilan.ac.ir}
\author{$^1$A. Ravanpak}
\author{$^3$H. Shojaie}
\author{$^1$M. Abolghasemi}
\affiliation{$^1$Department of Physics, University of Guilan, Rasht, Iran}
\affiliation{$^2$ School of Physics, University of New South Wales, Sydney, NSW, 2052, Australia}
\affiliation{$^3$Department of Physics, Shahid Beheshti University, G.C., Evin, Tehran 1983963113, Iran}

\def\be{\begin{equation}}
\def\ee{\end{equation}}
\def\bea{\begin{eqnarray}}
\def\eea{\end{eqnarray}}
\def\M{{{\cal M}}}
\def\bdy{{\partial\cal M}}
\def\w{\widehat}
\def\n{\widetilde}
\def\real{{\bf R}}

\begin{abstract}
This work is to study the generalized second law (GSL) of thermodynamics
in tachyon cosmology where the boundary of the universe is assumed to be enclosed by
a dynamical apparent horizon. The model is constrained with the observational data. The two logarithmic and power law corrected entropy is also discussed and conditions to validate the GSL and corrected entropies are obtained.
\end{abstract}

\pacs{}

\keywords{Tachyon cosmology; thermodynamics; GSL; entropy corrected; cosmic acceleration}
\maketitle
\newpage

\section{introduction}

Observations confirm that about two third of the content of the universe is filled with a component dubbed as dark energy (DE) which causes the cosmic expansion to accelerate~\cite{jarosik}. The cosmological constant is the simplest candidate for dark energy which fit the observational data. However, the measured expansion rate of the universe sets its scale being of order of $10^{-12}$GeV, which severely suffers from fine-tuning  when compared with the Planck scale ($10^{18}$GeV)~\cite{shaw}. Alternatively, there are a number of DE models which exploit scalar fields or some other exotic fields like phantom fields with negative energy~\cite{caldwell,piao}. However, such scalar fields are usually so light (or order of $10^{-33}$eV) which need an extremely fine tuning. It is also claimed in the literature, that the cosmic evolution of some kinds of these fields contradict with the solar system tests~\cite{nojiri}. It is most often the case that such fields interact with matter, i) directly due to a Lagrangian coupling, ii) indirectly through a coupling to the Ricci scalar or as the
result of quantum loop corrections~\cite{Damour, Carroll,Carroll2,bisvas}. Recently, a new case has been proposed in which a tachyon scalar field non-minimally couples to matter Lagrangian and the validity of this model in many scenarios has been investigated~\cite{Faraj1,Faraj2,Faraj3,Faraj4,Faraj5,Faraj6}.

On the other hand, inspired by the black hole physics, there is a deep connection between gravity and thermodynamics. An evidence for this connection in general relativity (GR) can be shown in deriving the Einstein equations in Rindler spacetime by using the proportionality of the entropy and the horizon area as well as the first law of thermodynamics. Moreover, the validity of the generalized second law of thermodynamics (GSL)~\cite{Davies,Pollock,Pavon1,Mukohyama,Brustein,Lineweaver,Izquierdo,Mohseni,Gong1,Horvat} has been under studies. According to GSL, the entropy of the fluid inside the horizon in addition to the entropy associated with the apparent horizon is a nondecreasing function of time. For instance, it is interesting that one can obtain the Friedmann equations by applying the Clasius relation to the apparent horizon of Friedmann-Robertson-Walker (FRW) universe~\cite{ca}. However, it should be mentioned that the definition of entropy would rather be modified in order to include quantum effects motivated from the loop quantum gravity~\cite{rov,ash}. The modification can be assumed to be a logarithmic or a power-law correction to entropy which consequently leads to appearance of correction terms in derived physical equations, such as modified Newtonian gravity, modified Friedmann equations~\cite{shey1,shey2,Deng}, and entropic corrections to Coulomb's law~\cite{shey3}.

In this paper, the GSL is studied in a general model, first in the standard entropy relation in Section II, and then in the corrected one in both logarithmic and power-law situations in Section III.  Then in Section IV, the universe in the presence of a tachyon scalar field non-minimally coupled to matter Lagrangian in the action is considered. The model is constrained with the observational data for distance modulus. In Section V, the GSL is explored by considering the logarithmic and power law corrections to the entropy and with the best-fit parameters.

\section{The GSL in a general model}

Considering the spatially flat Robertson-Walker (RW) metric, the Friedmann equations are
\bea
  3H^2 &=& 8\pi\rho \label{friedmann}\\
  2\dot H+3H^2 &=& -8\pi P \label{acceleration}.
\eea
However, in many cases dealing with a modified action of GR, one can rewrite these equations as
\bea
  3H^2 &=& 8\pi\rho_{eff} \label{friedmanneff}\\
  2\dot H+3H^2 &=& -8\pi P_{eff} \label{accelerationeff},
\eea
where $\rho_{eff}$ and $P_{eff}$ are effective density and pressure parameters respectively, which are related together by an effective equation of state parameter, namely $w_{eff}=P_{eff}/\rho_{eff}$.

In what follows, two assumptions are made: i) an entropy is associated with the horizon in addition to the entropy of the universe inside the horizon; ii) according to the local
equilibrium hypothesis, there is no spontaneous exchange of energy between the horizon and
the fluid inside.

GSL implies that in an expanding universe, the entropy of the viscous DE, dark matter and radiation
inside the horizon together with the entropy associated with the horizon do not decrease with
time. In general, there are two approaches to validate the GSL on the horizon: i)
by applying the first law of thermodynamics to find the entropy relation on the horizon~\cite{ca,bous}, i.e.,
\be\label{first law}
T_hdS_h=-dE_h=4\pi H R^3_h T_{\mu\nu}K^\mu K^\nu dt=4\pi (\rho_{eff}+P_{eff}) H R^3_h dt,
\ee
where the index $h$ stands for the horizon and $K^\mu=(1,-Hr, 0, 0)$ is the (approximate) Killing vector (the generator of the horizon),
or the future directed ingoing null vector field~\cite{wan}; ii)
in the field equations, by employing the horizon entropy and temperature relation on the
horizon~\cite{Bhattacharya},
\bea
S_h&=&\pi R^2_h \label{entropy}\\
T_h&=&\frac{1}{2\pi R_h} \label{temperature}.
\eea
Note that only on the apparent horizon are the two approaches equivalent \cite{Bhattacharya}. Furthermore,
the recent observational data from type Ia Supernovae suggests that in an accelerating
universe the enclosing surface would be the apparent horizon rather than the event one
~\cite{zh}. Therefore, the universe is assumed to be enclosed by a dynamical
apparent horizon with the radius~\cite{san}
\be\label{Rh}
 R_h=H^{-1}.
\ee
The equation~(\ref{entropy}) yields the dynamics
of the entropy as~\cite{re},
\be\label{dotSh}
\dot{S_h}=2\pi R_h\dot{R_h}.
\ee
Using~(\ref{Rh}) as well as the Friedmann equations~(\ref{friedmanneff},\ref{accelerationeff}) in~(\ref{dotSh}), leads to
\be\label{dotSheff}
\dot{S_{h}}=3\pi (1+w_{eff})R_h.
\ee
On the other hand, applying the fundamental thermodynamics relation to the fluid inside the horizon~\cite{go} gives
\be\label{TdSin}
TdS_{in}=P_{eff}dV+d E_{in}
\ee
where $S_{in}$ is the entropy within the apparent horizon, $P_{eff}$ is the effective pressure,
$E_{in}=\rho_{eff} V$ is the internal energy, and $V=\frac{4}{3}\pi R^3_h$. If there is no energy exchange between outside and inside of the apparent horizon, thermal equilibrium occurs that $T = T_{h}$. Therefore, in the case of thermal equilibrium, the relation~(\ref{TdSin}) gives
\be\label{dotSin}
\dot{S_{in}}=\frac{3}{2}\pi (1+w_{eff})(1+3w_{eff})R_h,
\ee
where we use Friedmann equations together with~(\ref{temperature}) and~(\ref{Rh}). Table~(\ref{t1}) summarizes dependencies of $\dot S_h$ and $\dot S_{in}$ on the equation of state.

\begin{table}
  \begin{center}
      \begin{tabular}{|c|c|c|c|c|c|}
      \hline
      & $w_{eff}<-1$ & $w_{eff}=-1$ & $-1<w_{eff}<-1/3$ & $w_{eff}=-1/3$ & $w_{eff}>-1/3$ \\ \hline
      $\dot S_{in}$ & + & 0 & -- & 0 & + \\
      $\dot S_h$ & -- & 0 & + & + & + \\
      \hline
      \end{tabular}
  \end{center}
  \caption{The behaviour of $\dot S_{in}$ and $\dot S_h$ with respect to $w_{eff}$}\label{t1}
\end{table}

From Table I, we observe that in cases where the universe is accelerating or crossing phantom line, the rate of change of the entropy inside the apparent horizon and on the apparent horizon is a decreasing function of time. On the other hand, the rate of change of the total entropy, namely the entropy of the horizon plus the entropy within the horizon is
\be\label{dotSt}
\dot{S_t}=\frac{9}{2}\pi (1+w_{eff})^2R_h,
\ee
which, regardless of the behavior of the apparent horizon radius, is obviously a nondecreasing function of time. The relations~(\ref{dotSheff}),~(\ref{dotSin}) and~(\ref{dotSt}) can be re-parameterized
as
\be\label{Sz}
\frac{dS_X}{dz}=-\frac{R_h}{1+z}\dot S_X,
\ee
where $z$ is the redshift and the index $X$ stands for the indices ``$in$'', ``$h$'' and ``$t$'' respectively. As it is apparent from~(\ref{Sz}), in an ever-expanding universe, $dS_X/dz\lessgtr0$ whenever $\dot S_X\gtrless0$.

\section{CORRECTIONS TO ENTROPY}

As mentioned in standard general theory of relativity, the horizon entropy is proportional to the area of the horizon, i.e. $S=A/4$ where $A=4\pi R_h^2$. However, modification of the theory, due to the motivation from loop quantum gravity, leads to a correction to the above relation. For instance, in $f(R)$ gravity, the modified entropy is $f'(R)A$~\cite{Wald}. Quantum corrections to the semi-classical entropy law, on the other hand, have been often introduced by logarithmic and power-law terms. In the following contributions from these kinds of corrections will be discussed.

\subsection{Logarithmic corrections}

Logarithmic corrections, arises from loop quantum gravity due to the thermal equilibrium and quantum fluctuations \cite{Miessner,Ghosh,Chatterjee,Banerjee,Modak,Jamil,Sadjadi} and as a result the entropy on the apparent horizon become,
\be\label{sh1}
S_{h}=\frac{A}{4}+\alpha \ln \frac{A}{4}
\ee
where $\alpha$ is a dimensionless constant of order unity whose value is still matter of debate. Following the same procedures resulting the relation~(\ref{dotSheff}) and~(\ref{dotSt}), one finds the apparent horizon and total entropy as
\be\label{dsh1}
\dot{S_h}=3(1+w_{eff})(\pi R_h+\frac{\alpha}{R_h})
\ee
and
\be\label{dst1}
\dot{S_t}=\frac{9}{2}\pi(1+w_{eff})^2R_h+3 \alpha(1+w_{eff})R^{-1}_h.
\ee
respectively. According to~(\ref{dst1}), the GSL is satisfied if
\be\label{alfa1}
\alpha\geq\beta\equiv-\frac{3}{2}\pi(1+w_{eff})R^2_h.
\ee

\subsection{Power-law corrections}

When dealing with the entanglement of quantum fields in
and out the horizon, power-law corrections appear, as for example~\cite{Radicella,Sheykhi,Farooq,Das}
\be\label{sh2}
S_{h}=\frac{A}{4}\left(1-K_\alpha A^{1-\frac{\alpha}{2}}\right),
\ee
where
\be\label{kalfa}
K_{\alpha}=\frac{\alpha (4 \pi)^{\frac{\alpha}{2}-1}}{(4-\alpha)r_{c}^{2-\alpha}}
\ee
and $r_c$ is the crossover scale. The second term in~(\ref{sh2}), as a power-law correction to the entropy, has been raised from the entanglement of the wave-functions of a scalar field between its ground state
and an exited state. The higher the excitation state the more significant the correction term. It
is worth noting that in~(\ref{kalfa}), the correction term tends to zero as the semi-classical
limit (large area) is retrieved and consequently the conventional entropy is recovered. The rates of change of the horizon entropy and total entropy are
\be\label{dsh2}
\dot{S_h}=3 \pi (1+w_{eff})\left(1-K_{\alpha}(4 \pi R^2_h)^{1-(2-\frac{\alpha}{2})\frac{\alpha}{2}}\right)R_h
\ee
and
\be\label{dst2}
\dot{S_t}=\frac{9}{2} \pi (1+w_{eff})^2R_h-\frac{3}{2}\pi(1+w_{eff})\alpha \frac{R_h^{3-\alpha}}{r_c^{2-\alpha}}
\ee
respectively. Therefore, the GSL will be valid if
\be\label{alfa2}
\alpha\left(\frac{r_c}{R_h}\right)^\alpha\leq 3(1+w_{eff})\left(\frac{r_c}{R_h}\right)^2.
\ee
Following \cite{Radicella,Dvali,Karami,Hendi}, by replacing $r_c$ with $R_{h0}$, the above constraint becomes
\begin{equation}\label{alfa3}
\alpha\left(\frac{R_{h0}}{R_h}\right)^\alpha\leq 3(1+w_{eff})\left(\frac{R_{h0}}{R_h}\right)^2.
\end{equation}
In the next section, these two corrected version of entropy will be employed to study GSL in the tachyon cosmology in the presence of a non-minimal coupling term to matter.

\section{Tachyon Cosmology}

The tachyon cosmology with a non-minimal coupling to matter is given by the
action:
\be\label{action}
S=\int{d^4x\sqrt{-g}\left(\frac{R}{16\pi}-V(\phi)\sqrt{1+g^{\mu\nu}\partial_\mu\phi\partial_\nu\phi}+f(\phi){\cal{L}}_m\right)}
\ee
where $R$ is Ricci scalar and $V(\phi)$ denotes the tachyon potential. Unlike the usual Einstein-Hilbert action, the matter Lagrangian ${\cal{L}}_m$ is modified as $f(\phi){\cal{L}}_m$, where $f(\phi)$ is an analytic function of $\phi$ and causes a non-minimal coupling between the matter and the scalar field. We assume that the mater field filled the universe is cold dark matter and also a spatially flat FRW metric, which forces the tachyon scalar field to be a function of only cosmic time, the field equations become
\be\label{friedmann1}
3H^2=8\pi\left(\rho_m f(\phi)+\frac{V(\phi)}{\sqrt{1-\dot{\phi}^2}}\right),
\ee
\be\label{friedmann2}
2\dot{H}+3H^2=-8\pi\left(-V(\phi)\sqrt{1-\dot\phi^2}\right)
\ee
and
\be\label{scaler field}
\ddot{\phi}+(1-\dot{\phi}^2)\left(3H\dot{\phi}+\frac{V^\prime(\phi)}{V(\phi)}\right)=\frac{f^\prime(\phi)}{4V(\phi)}(1-\dot{\phi}^2)^\frac{3}{2}\rho_m
\ee
where \emph{prime} denotes differentiation with respect to the tachyon field $\phi$.

By using~(\ref{friedmann1}) and~(\ref{scaler field}), One can easily arrive at the generalized conservation equation:
\be\label{conservation}
\dot{(\rho_mf(\phi))}+3H\rho_mf(\phi)=-1/4 \rho_m\dot{f}(\phi).
\ee
In comparison with~(\ref{friedmanneff}) and~(\ref{accelerationeff}), the effective equation of state parameter is identified
\begin{equation}\label{EoS}
w_{eff}=\frac{-V(\phi)\sqrt{1-\dot{\phi}^2}}{\rho_mf(\phi)+\frac{V(\phi)}{\sqrt{1-\dot{\phi}^2}}}.
\end{equation}
In the following we constrain the parameters of model by using the observational data. Without losing the generality we first rewrite the equations by introducing the following new variables,
\be\label{variable}
X = \frac{8\pi \rho_mf}{3H^2} , \quad Y = \frac{8\pi V}{3H^2} , \quad Z = \dot{\phi}, \quad U = \frac{1}{H}.
\ee
We assume that $f(\phi)=f_0 \exp(\delta_1\phi)$ and $V(\phi)=V_0 \exp(\delta_2 \phi)$ where $\delta_1$ and $\delta_2$ are dimensionless
constants characterizing the slope of the potential $V(\phi)$ and the coupling field $f(\phi)$. Cosmological model with the potentials in the form of exponential functions leading to interesting
physics have been used in a variety of contexts, such as accelerating expansion cosmological models~\cite{do},
cosmological scaling solutions~\cite{se,chahar,a,b}, chameleon cosmological models~\cite{noh,c}, and models with attractor solutions ~\cite{panj,d,e} ~\cite{shesh,f,g}.
Using~(\ref{friedmann1})-(\ref{conservation}), the equations for the new dynamical variables become
\be\label{variable1}
\frac{dX}{dN}=X\left(-1/4\delta_1ZU-3Y\sqrt{1-Z^2}\right)
\ee
\be\label{variable2}
\frac{dY}{dN}=Y\left(\delta_2ZU+3-3Y\sqrt{1-Z^2}\right)
\ee
\be\label{variable3}
\frac{dZ}{dN}= \frac{1/4\delta_1XU(1-Z^2)^\frac{3}{2}}{Y}-(1-Z^2)(3Z+\delta_2U)
\ee
\be\label{variable4}
\frac{dU}{dN}=\frac{3}{2}U\left(1-Y\sqrt{1-Z^2}\right),
\ee
where $N=\ln a$ with $a$ being the scalar factor. The Friedmann equation~(\ref{friedmann1}), in terms of the new dynamical variables is
\be\label{variable5}
X+\frac{Y}{\sqrt{1-Z^2}}=1.
\ee
Using the Friedmann constraint, equations~(\ref{variable1})-(\ref{variable4}) reduce to
\be\label{variable6}
\frac{dX}{dN}=X\left(-1/4\delta_1ZU-3(1-X)(1-Z^2)\right)
\ee
\be\label{variable7}
\frac{dZ}{dN}=\frac{1/4\delta_1XU(1-Z^2)}{1-X}-(1-Z^2)(3Z+\delta_2U)
\ee
\be\label{variable8}
\frac{dU}{dN}=\frac{3}{2}U\left(1-(1-X)(1-Z^2)\right).
\ee
Finally, the effective equation of state parameter, as a function of these new variables, is written as
\begin{equation}\label{weffnew}
w_{eff}=\frac{-Y(1-Z^2)}{X\sqrt{1-Z^2}+Y}.
\end{equation}
To constrain the model parameters with the recent observational data, we use ``Union2'' sample~\cite{Aman} consisting of 557 usable SNe Ia data. The $\chi^2$ method, is the method used to best-fit the model for the parameters $\delta_{1}$ and $\delta_{2}$, and the initial conditions $X(0)$, $Z(0)$ and $U(0)$. In this case, the $\chi^2$ function is introduced as
\be\label{sne2}
\chi_{SNe}^2(\delta_{1},\delta_{2}, X(0),Z(0),U(0))=\sum_{i=1}^{557}\frac{\left[\mu_i^{the}(z_i|\delta_{1},\delta_2, X(0),Z(0),U(0))-\mu_i^{obs}\right]^2}{\sigma_i^2},
\ee
where $\mu_{i}^{the}$ and $\mu_{i}^{obs}$ are the distance modulus parameters obtained from our model and the observations, respectively, and $\sigma$ is the estimated error of $\mu_i^{obs}$. It should be mentioned that the difference between the absolute and the apparent luminosity of a distance object is given by
\be\label{Ab-Ap}
\mu(z)=5 \log_{10}D_{L} (z)-\mu_0
\ee
where $\mu_0=5\log_{10}h-42.38$ and $h=(H_0/100)$km/s/Mpc. The Luminosity distance quantity, $D_{L}(z)$ in~(\ref{Ab-Ap}) is derived as
\be\label{sne}
D_{L} (z)=(1+z)H_{0} \int_{0}^z \frac{dz^\prime}{H(z^\prime)}.
\ee
Table~\ref{t2}, summarizes the results derived by minimizing~(\ref{sne2}).

\begin{table}
 \begin{center}
    \begin{tabular}{|c|c|c|c|c|c|}
    \hline
    Model parameters  &  X(0) & Z(0) & U(0) & $\delta_{1}$ & $\delta_{2}$ \\ \hline
    Best-fitting results & 0.14 & 0.38 & 1.43 & -0.9 & -0.5 \\
    \hline
    \end{tabular}
 \end{center}
 \caption{Best-fit values}\label{t2}
\end{table}

\section{GSL}

\begin{figure}[h]
    \centering
    \includegraphics[width=0.32\textwidth]{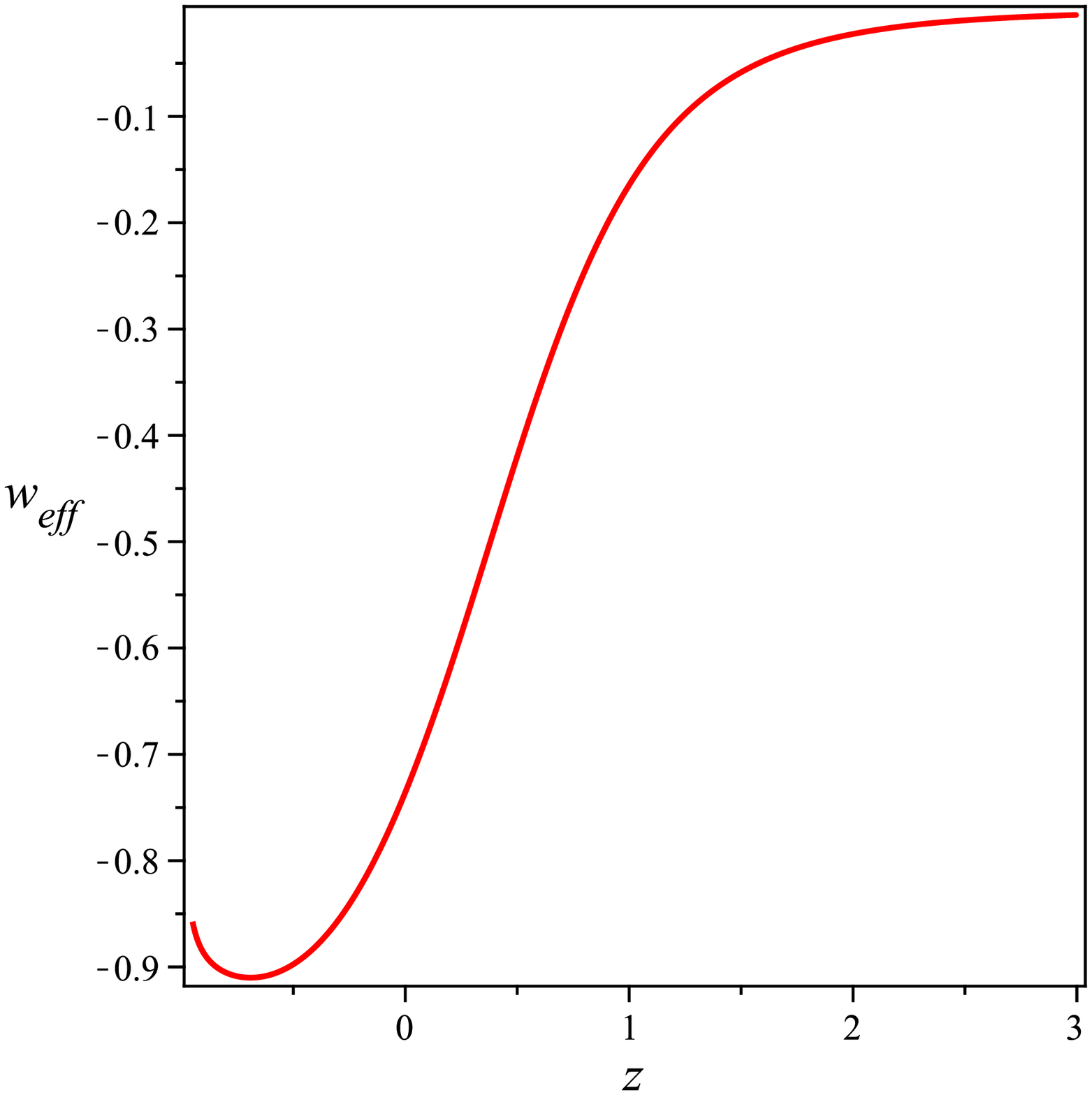}
    \includegraphics[width=0.32\textwidth]{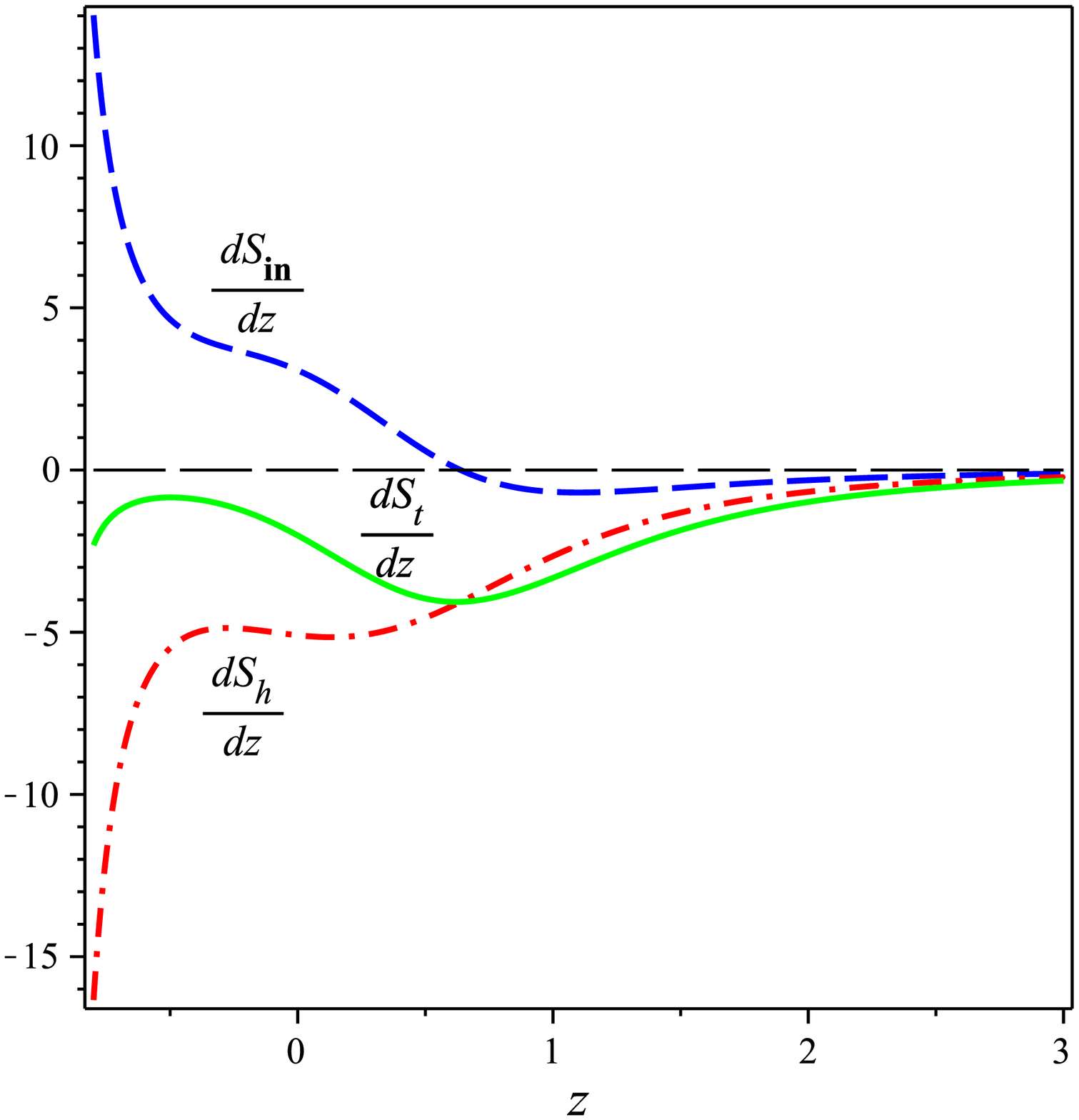}
    \includegraphics[width=0.32\textwidth]{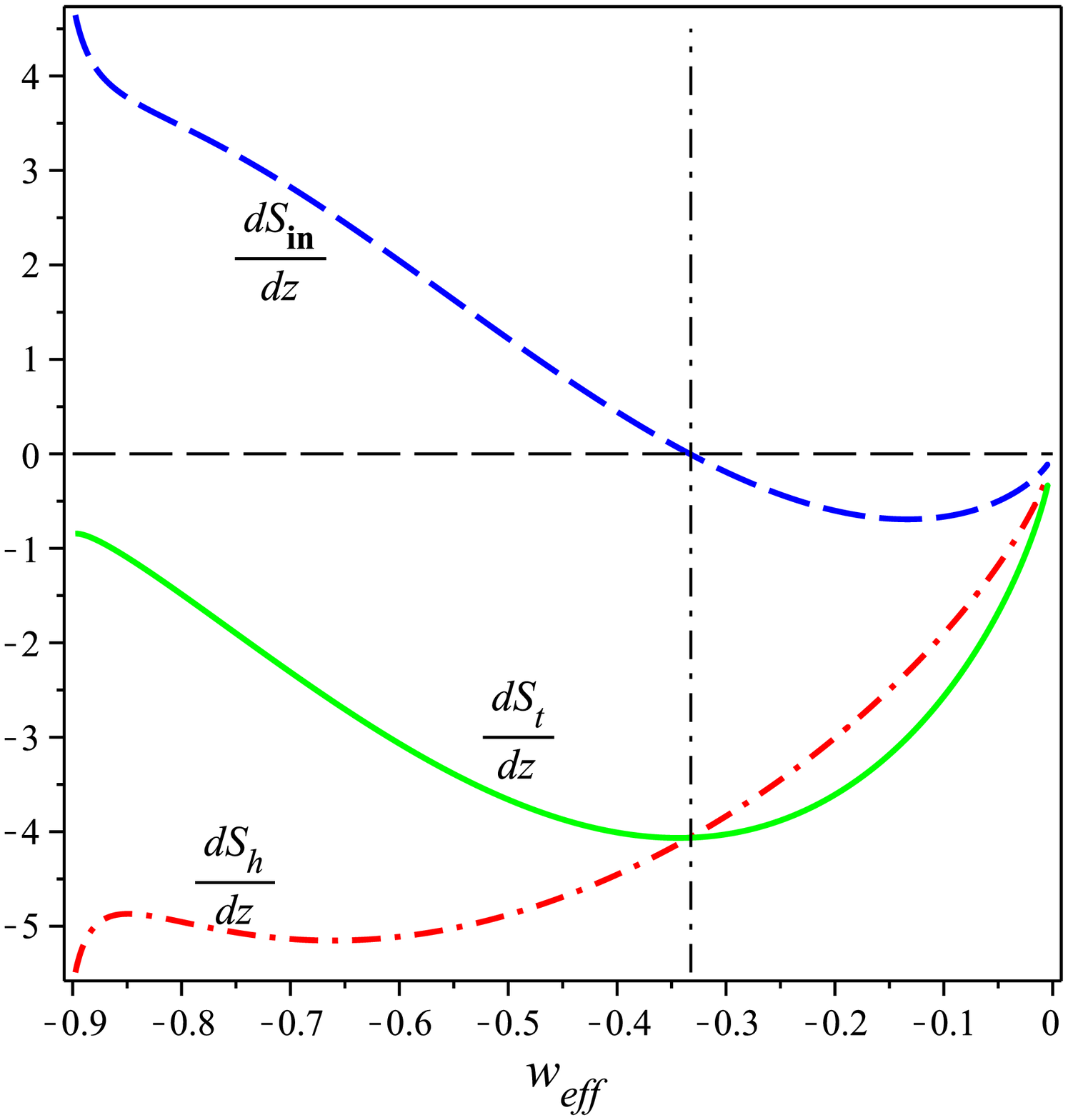}
    \caption{Left: the dynamics of $w_{eff}$ vs. the redshift. Middle: the dynamics of $\frac{dS_h}{dz}$, $\frac{dS_{in}}{dz}$ and $\frac{dS_t}{dz}$ vs. the redshift. Right: The dynamics of $\frac{dS_h}{dz}$, $\frac{dS_{in}}{dz}$ and $\frac{dS_t}{dz}$ vs. $w_{eff}$.}\label{S}
\end{figure}

In this section, by using the best fitted model parameters, the GSL is investigated. Fig.~\ref{S} is an illustration of cosmological parameters such as effective EoS parameter in relation to the rate of change of entropy. The graphs are for out model in hand when there is no correction to the entropy. The left panel in Fig.~\ref{S}, is the graph of $w_{eff}$ versus the redshift $z$. The plot shows that $w_{eff}>-1$ in the;past and future; the universe is currently in quintessence era and it never crosses the phantom barrier. To probe GSL, using~(\ref{dotSh}),~(\ref{dotSin}) and~(\ref{dotSt}), we plot $dS_h/dz$, $dS_{in}/dz$ and $dS_t/dz$ versus the redshift (the middle panel), and also versus $w_{eff}$ (the right panel) of Fig.~\ref{S}. The middle graph in Fig.~\ref{S} shows the rate of change of entropy within the apparent horizon, on the apparent horizon and total.

\begin{figure}[h]
\centering
\includegraphics[width=0.42\textwidth]{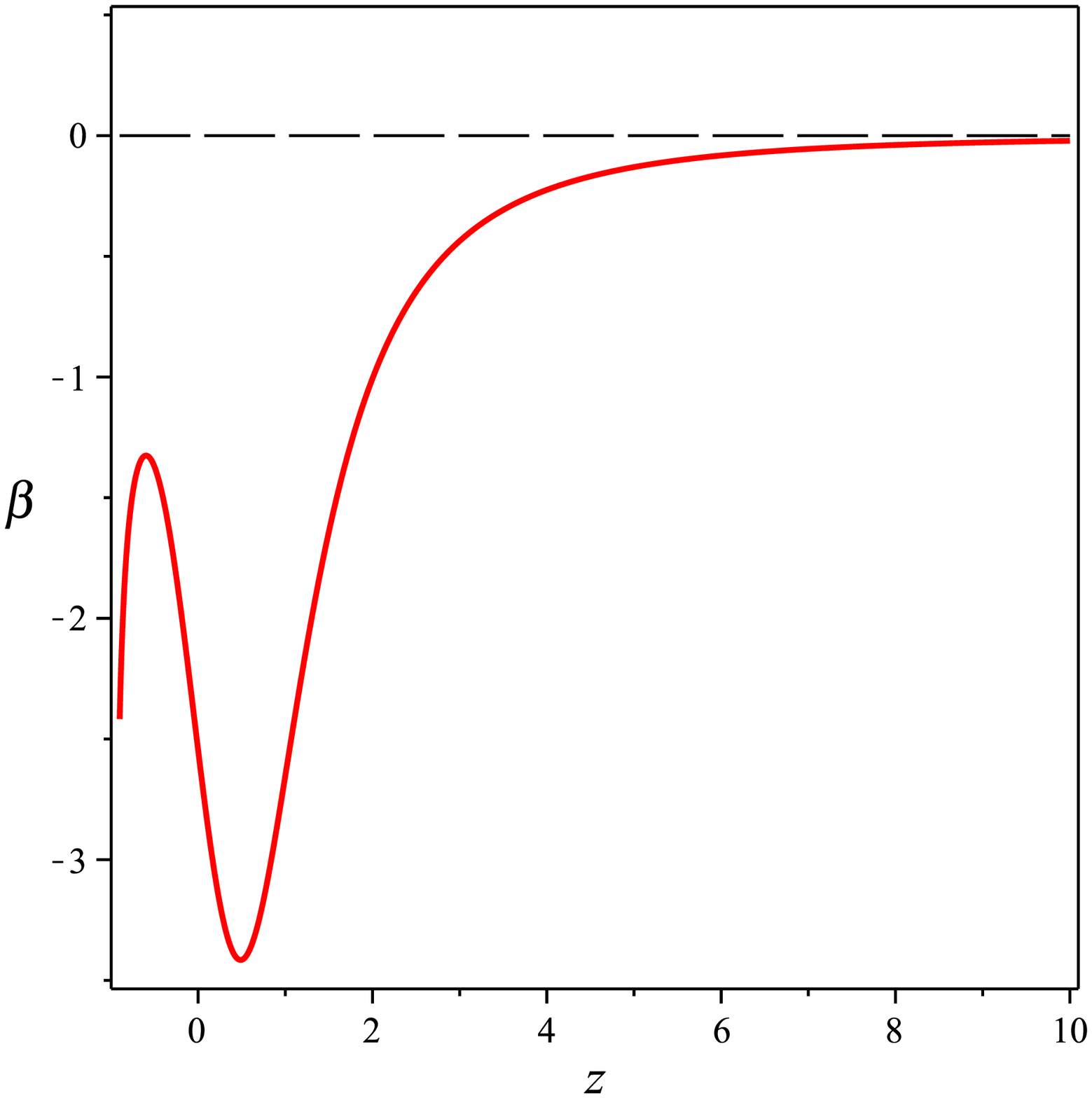}
\includegraphics[width=0.42\textwidth]{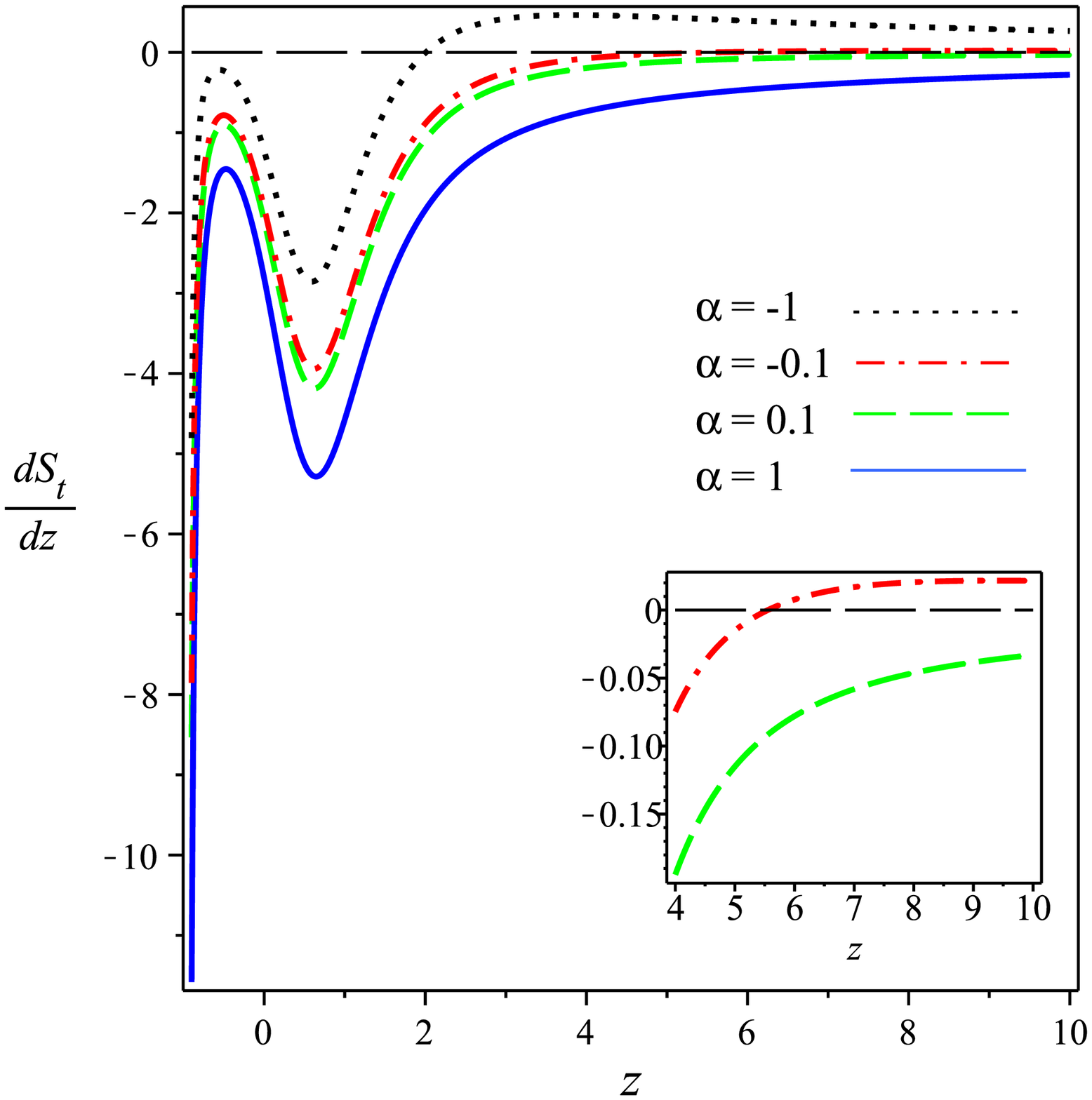}
\caption{Left: the dynamics of $\beta$ vs. the redshift., and right: the dynamics of $\frac{dS_t}{dz}$ vs. the redshift, in logarithmic corrected case for different values of $\alpha$.}\label{alpha}
\end{figure}

In the case of logarithmic corrections, we plot $\beta$, the right hand side of~(\ref{alfa1}), in the left panel of Fig.~\ref{alpha}. The constant parameter $\beta$ approaches zero from below. To satisfy GSL, this forces the constant parameter $\alpha$ to be nonnegative as it is apparent from ~(\ref{alfa1}). The right panel of Fig.~\ref{alpha} is an evidence for this claim.

\begin{figure}[ht]
\centering
\includegraphics[width=0.42\textwidth]{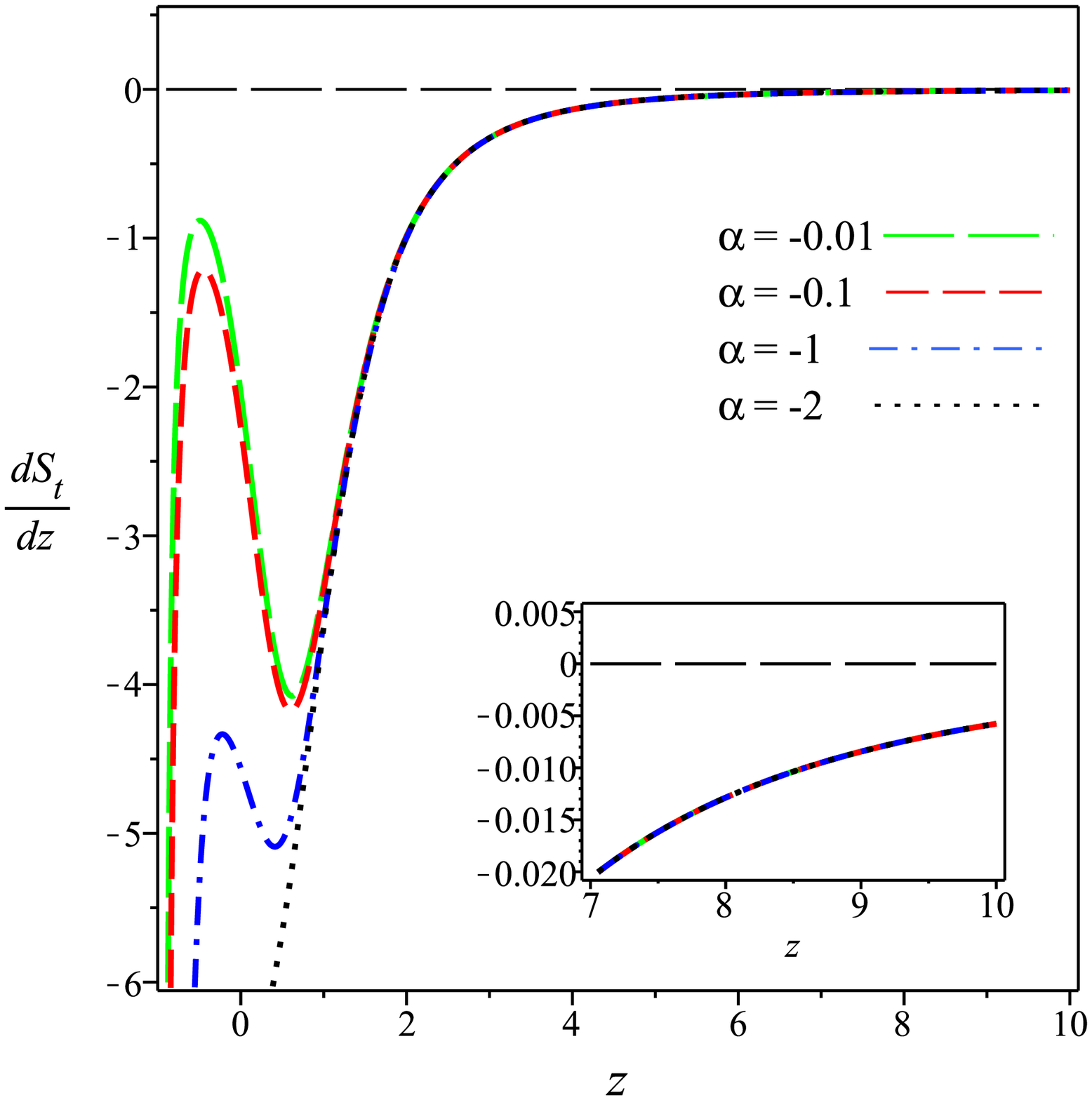}
\includegraphics[width=0.42\textwidth]{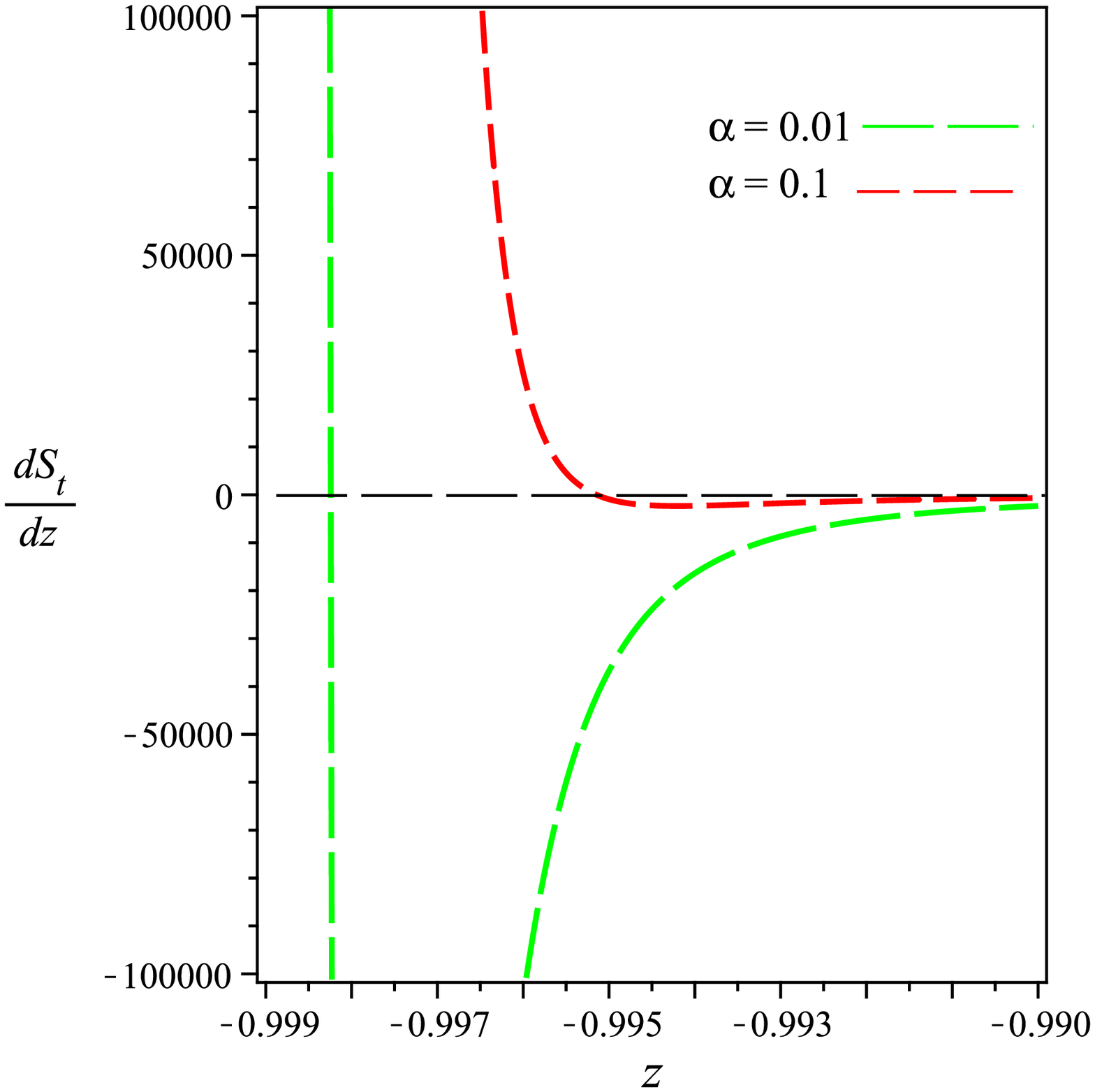}
\includegraphics[width=0.42\textwidth]{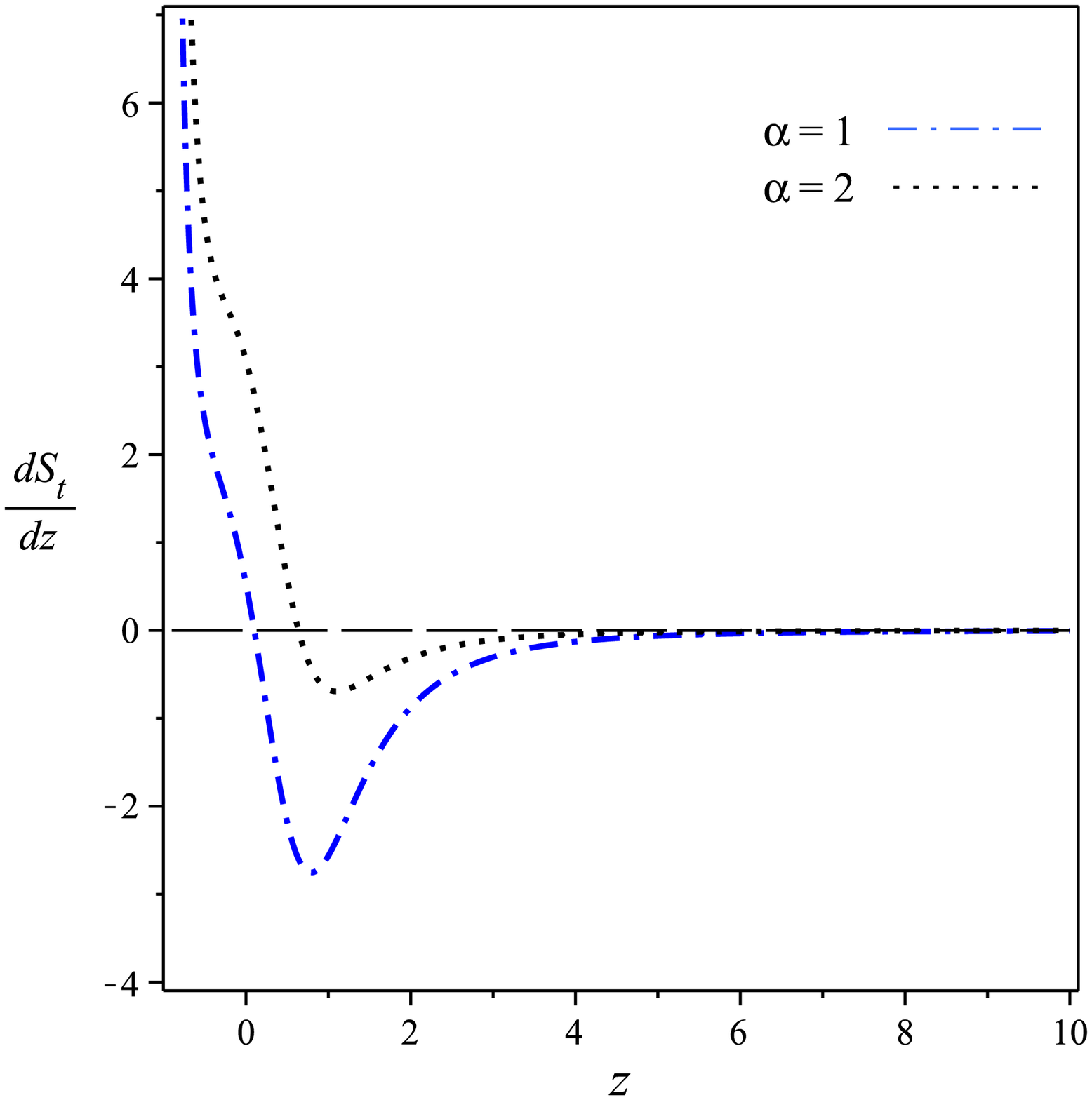}
\caption{The dynamics of $\frac{dS_t}{dz}$ vs. the redshift in power-law corrected case different values of $\alpha\leq0.1$.}\label{alpha2}
\end{figure}

Finally, in the case of power-law entropy corrections, we plot $\frac{dS_t}{dz}$ against redshift for different values of parameter $\alpha$ in~(\ref{dst2}). The graphs in Fig.~\ref{alpha2} confirm that the GSL is satisfied for $\alpha\leq 0$ . But as $\alpha$ becomes positive, the GSL is violated in the past or future.

\section{Summary and remarks}

This paper is intended to study GSL and its corrected version in tachyon cosmological model in the presence of a non-minimal coupling to the matter field. The model is best fitted with the observational data in order to validate the findings. The entropy for the apparent horizon, the fluid inside the universe and the total entropy is obtained. To include quantum effects motivated from loop quantum gravity, two kinds of corrections to the entropy; logarithmic and power law corrections are investigated. In the logarithmic correction, the GSL is satisfied for $\alpha \geq 0$ and in power law for $\alpha \leq 0$. Note that since the best fitted model with observational data reveals no phantom crossing in the past and future the rate of change of entropy of apparent horizon, $dS_{h}/dz$, does not change the behavior and is always positive in accelerating and decelerating eras. On the other hand, the universe begins to accelerate at about $z\simeq -0.6$, and thus the entropy within the horison, $dS_{in}/dz$, changes from negative ( decelerating era) to positive (accelerating era). However,under any circumstances the total entropy variation, $dS_{t}/dz$, is always negative in both cases.

\section{Acknowledgment}

We would like to thank the university research council of the University of Guilan for financial support.

\end{document}